\documentclass[a4paper,fleqn,usenatbib]{mnras}

\usepackage{newtxtext,newtxmath}

\usepackage[T1]{fontenc}
\usepackage{ae,aecompl}

\usepackage{graphicx}	
\usepackage{amsmath}	
\usepackage{amssymb}	
\graphicspath{ {images/} }
\usepackage{natbib}
\usepackage{tablefootnote}
\usepackage{float}

\title[An Archival \textit{XMM-Newton} Observation of XY Ari]{A Long Serendipitous \textit{XMM-Newton} Observation of the Intermediate Polar XY Ari}

\author[Zengin \c{C}amurdan et al.]{
D. Zengin \c{C}amurdan,$^{1,3}$\thanks{E-mail: dicle.zengincamurdan@ege.edu.tr}
\c{S}. Balman,$^{2}$\thanks{solen@astroa.physics.metu.edu.tr}
V. Burwitz $^{3}$\thanks{burwitz@mpe.mpg.de}
\\
$^{1}$Department of Astronomy and Space Sciences, Ege University, 35100 Bornova, \.{I}zmir, Turkey \\
$^{2}$Middle East Technical University, Dept. of Physics, Dumlup{\i}nar Bulvar{\i} Universiteler Mah. No.1, 06800, Ankara,Turkey\\
$^{3}$Max-Planck-Institut f{\"u}r extraterrestrische Physik, Giessenbachstrasse 1, D-85748, Garching, Germany
}

\date{Accepted XXX. Received YYY; in original form ZZZ}

\pubyear{2018}

\begin{document}
\label{firstpage}
\pagerange{\pageref{firstpage}--\pageref{lastpage}}
\maketitle

\begin{abstract}
XY Ari is one of the few known eclipsing intermediate polars. We present results from a detailed analysis of an unpublished archival observation using \textit{XMM-Newton} EPIC pn and MOS data in a quiescent state of XY Ari. The X-ray orbital modulation and spin pulse variations were investigated for energy dependent modulations in different energy bands. The broad orbital modulation observed with various observations was confirmed with \textit{XMM-Newton} at hard X-ray (>1.6 keV). The EPIC light curves folded at the spin phases show a double peak profile as expected from two pole accretion. The pulse profile is found to be energy dependent. Hardness ratio variations and energy modulation depth during spin modulation can be explained by photoelectric absorption. The simultaneously fitted EPIC spectra with CEVMKL model yield maximum plasma temperature of $28^{+3.1}_{-2.9}$ keV with an iron abundance $\mathrm{Fe}/\mathrm{Fe}_\odot=0.37^{+0.06}_{-0.05}$. We find two intrinsic partial covering absorption columns of $6.2^{+1.0}_{-0.9} \times 10^{22}$ and $105.3^{+35.4}_{-30.4} \times 10^{22} \,\mathrm{cm^{-2}}$ with covering fractions of $0.53^{+0.05}_{-0.04}$, $0.41^{+0.14}_{-0.13}$ respectively. In addition, a Gaussian emission line at $6.43^{+0.01}_{-0.02}$ keV with an equivalent width of $51^{+12}_{-10}$ eV is required to account for fluorescent emission from neutral iron. The X-ray luminosity of the source is $4.2 \times 10^{32} \,\mathrm{erg \,s^{-1}}$ in the 0.2-10.0 keV energy band. 

\end{abstract}




\section{Introduction}

Cataclysmic Variables (CVs) are close binary star systems consisting of a white dwarf (WD) and a low mass main-sequence star. When the binary system evolves, the secondary fills up the Roche-lobe and begins to transfer material onto the white dwarf surface due to its strong gravitational field \citep{Warner95}. This mass accretion releases large amounts of gravitational energy which makes CVs powerful emitters of radiation from infrared to X-ray wavelengths and causes variations that are either random or periodic. The magnetic field of the white dwarf influences the accretion flow from its companion and hence magnetic CVs (MCVs) have two distinct subclasses, polars (or AM Her type systems) and intermediate polars (or DQ Her type systems). No accretion disk can form around polar MCVs where the white dwarf is strongly magnetic (10-230 MG) \citep{Beuermann88,Crop90}. The accretion is magnetically funnelled directly onto the magnetic poles of the white dwarf. The spin period of white dwarf is generally synchronized to the orbital period. In the intermediate polars (IP) where the white dwarf primary is weakly magnetic (0.1-10 MG) an accretion disk is expected to exist but the inner disk is truncated depending on the strength of the magnetic field. The material is forced to flow along the field lines. The accretion occurs through a truncated disk and via accretion curtains to the magnetic poles of the white dwarf \citep[see][]{Rosen88,Pat94}. In these systems the WD rotates asynchronously with its companion ($P_{\mathrm{spin}} \ll P_{\mathrm{orb}}$). In addition, accretion disk outbursts are sometimes observed.

XY Ari which belongs to the group of IPs, was discovered as an X-ray source in the direction of molecular cloud MBM 12 (also known as Lynds 1457) by the \textit{Einstein} satellite \citep{Halp87}. The detection of the coherent X-ray pulsations by \textit{Ginga} satellite at a period of 206 s revealed the magnetic nature of cataclysmic variable XY Ari \citep{Takano89,Koyama91}. The folded pulse profiles were found to be double-peaked indicating two emitting poles. \cite{Kamata91} discovered deep X-ray dips with a 6.06 h orbital period  and determined the eclipse duration of $1990\pm30$ s using the observations with the same satellite. The observations in the infrared revealed the ellipsoidal light variations of XY Ari that allowed the determination of some of the system parameters such as the mass ratio (0.43 $<$ q $<$ 0.71) and the orbital inclination ($i$ $<$ 84\degr) \citep{Hellier97}. The optical counterpart can not be observed because of the visual extinction in the direction of XY Ari is 11.5$\pm$0.3 mag \citep{Littlefair01}. The distance of the system was estimated as 270$\pm$100 pc  using \textit{K}-band photometry in the same study.

XY Ari is the only known IP that is bright in X-rays and displays an eclipsing X-ray light curve. EX Hya, which has an orbital period of 98 min, is another well-studied IP shows partial X-ray eclipses \citep{BuerOs88,Pekon11}. Compared to the mean X-ray luminosity of IPs, the X-ray luminosity of XY Ari ($\sim 3 \times 10^{32}  \, \mathrm{erg \,s^{-1}}$, \citealt{Kamata93}\footnote{The luminosity given in \cite{Kamata93} has been corrected using the updated distance of 270 pc \citep{Littlefair01}}, \citealt{Salinas04}) is about a factor of five higher than EX Hya ($\sim 6 \times 10^{31} \, \mathrm{erg \,s^{-1}}$, \citealt{Pekon11}). \cite{Hellier97} studied the size and the structure of X-ray emitting accretion region using a series of eclipse egresses observed with \textit{RXTE} satellite. In this study, the emitting area at each pole cap was found to have a size less than 0.002 of the white dwarf surface area. The only known outburst of XY Ari was observed by \citet*{HelMukBea97} in the same \textit{RXTE} observations. They also reported that during the outburst, the spin pulse evolved from a double peaked to a single peaked profile. The white dwarf mass has been estimated to be in the range 0.86-1.28 $M_{\sun}$ using infrared \citep{Allan96} and X-ray observations \citep{Hellier97,Ramsay98,Yuasa10}. The X-ray spectrum of the post-shock region of XY Ari has been fitted using an absorbed thermal Bremsstrahlung with $kT$ $\sim$ 16 - 30 keV in the previous studies \citep{Kamata93,Hellier97,Ramsay98,Salinas04,Norton07}. The three lines of Fe near 6.4, 6.7 and 6.9 keV were detected in the spectral studies \citep{Ezuka99,Salinas04,Yuasa10}. The emission lines reported in the spectrum of XY Ari by \cite{Salinas04} indicate multi-temperature nature of the shock-heated plasma emission.

A broad sinusoidal X-ray orbital modulation of XY Ari was observed in previous observations with \textit{Ginga} \citep{Kamata93} and \textit{Chandra} \citep{Salinas04}. \cite{Norton07} investigated the cause of this long term behaviour in detail using various observations. They demonstrated that the broad orbital modulation drifts in phase and in modulation depth. In addition, the X-ray spin pulse profile shows variations in depth  and in shape between double and single peaked pulses. They suggested a precessing, tilted accretion disc in the system may cause both observed changes in the broad orbital modulation and in the pulse profiles. In the recent observation which was acquired in 2006 by \textit{RXTE}, the broad orbital modulation couldn't be detected in their study giving the motivation of further study of XY Ari with more recent data to us. The observed changing behaviour of the modulation in different observations over a long time span and the variations in X-ray pulse profiles in depth and shape also motivate this work in order to study the accretion flow in this system. Additionally, the variation of absorption column over orbital phase was detected by \cite{Kamata93}. \cite{Salinas04} confirmed the variation and also reported spin-phase dependence of absorption. 

In this work we report on our analysis of the \textit{XMM-Newton} observation of XY Ari. This paper is targeted at reaching a greater understanding of the interplay between the X-ray emission sites and absorption that produces the observed modulation of the X-ray light curves as well as the spectral properties.  In Section 2, we describe the details of the X-ray observation as well as the procedure of the data reduction. The results of the orbital and the spin pulse profile analysis are presented in Sect. 3 and 4. The phase averaged spectrum and the pulse-phase spectroscopy are investigated in Sect. 5 and 6, respectively. Finally, we discuss and summarise our results in Sect. 7.

\begin{figure}
	\includegraphics[angle=90, width=\columnwidth]{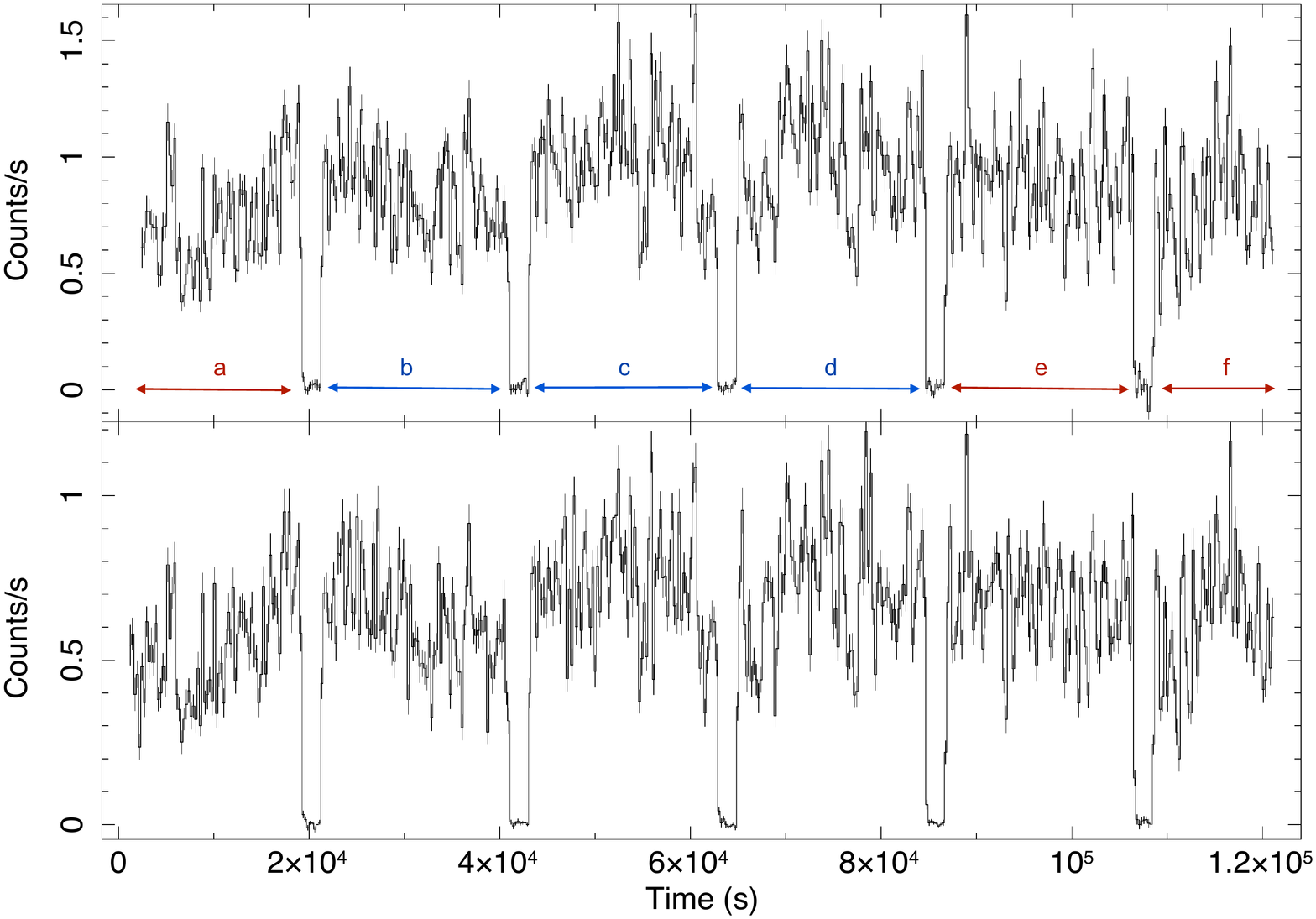}
    \caption{The 0.2-10 keV background subtracted light curves of XY Ari obtained from EPIC pn (top) and combined MOS (bottom) data. The letters a to f represent the data sets indicated by the blue and red arrows and explained in detail in Sec. [4.0]. The zero points of time scale for pn and MOS are TJD 15408.112 and 15408.097, respectively. The binning time is 100 s.}
    \label{Fig.1 }
\end{figure}

\section{Observation And Data  Reduction}

An archival observation of XY Ari obtained with the \textit{XMM-Newton} observatory was used in this study. It was observed for approximately 125 ks ($\sim$ 35 h) during quiescence state on July 31, 2010 (observation ID 0652480201), covering 5.95 cycles of the $\sim$ 6.06 h binary period. The EPIC pn \citep{Struder01} and MOS1 and MOS2 cameras \citep{Turner01} were operated in full frame imaging mode with the thin filter. 

We employed the \textit{XMM-Newton} software analysis system {\small SAS} v14.0.0 and the tasks for the data reduction. Since the exposure was affected by the flaring episodes, we flagged and removed the last 5.4 ks of the observation. For both EPIC instruments, we extracted the light curves and spectra from circular source and background regions with a radius of 25 arcsec. We used offset background regions on the same CCD chip for the background subtraction. We estimated the effects of the pile-up on all of the instruments by using \textit{epaplot} tool.

The EPIC pn and MOS data were processed with the standard tasks \textit{epchain} and \textit{emchain} to generate the calibrated event lists. For extracting the light curves, the spectrum of the source and the background, data with single and double-pixel events (PATTERN 0-4) for EPIC pn and single to quadruple events for EPIC MOS were extracted in 0.2-10 keV energy range using \textit{evselect} tool. We removed the bad CCD pixels and columns (FLAG = 0). The extraction of the background subtracted source spectrum was carried out using \textit{especget} tool. The average net count rates are 0.930$\pm$0.003 ct/s and 0.307$\pm$0.002 ct/s for the EPIC pn and the combined MOS1 and MOS2 data in the 0.2-10 keV energy interval, respectively. Even though the EPIC pn count rate is higher than the mean rates obtained previously in 2000 and 2001 observations ($0.877\pm0.007$, $0.786\pm0.005$, respectively)  by \cite{Anzo10}, these count rates confirm the quiescent state of XY Ari. Background-subtracted light curves were generated using \textit{lcmath} tool within {\small FTOOLS} (version 6.10). Photon arrival times were corrected to the barycenter of the solar system with the \textit{barycen} task. 

\section{The X-ray Light Curves of XY Ari}

As a result of the high inclination of the system \citep[$80\degr < i < 84\degr$,][]{Hellier97} we observe XY Ari nearly edge on and detect deep X-ray eclipses. \cite{Kamata91} argued that the deep X-ray eclipses arise from an eclipse of the white dwarf by the companion star. \cite{Hellier97} also assumed this scenario in order to study the size of the X-ray emitting accretion region. In this case, the secondary star occults the X-ray emitting region which covers a small area over WD (e.g. polar cap region).

 \begin{figure}
	\includegraphics[angle=0, width=\columnwidth]{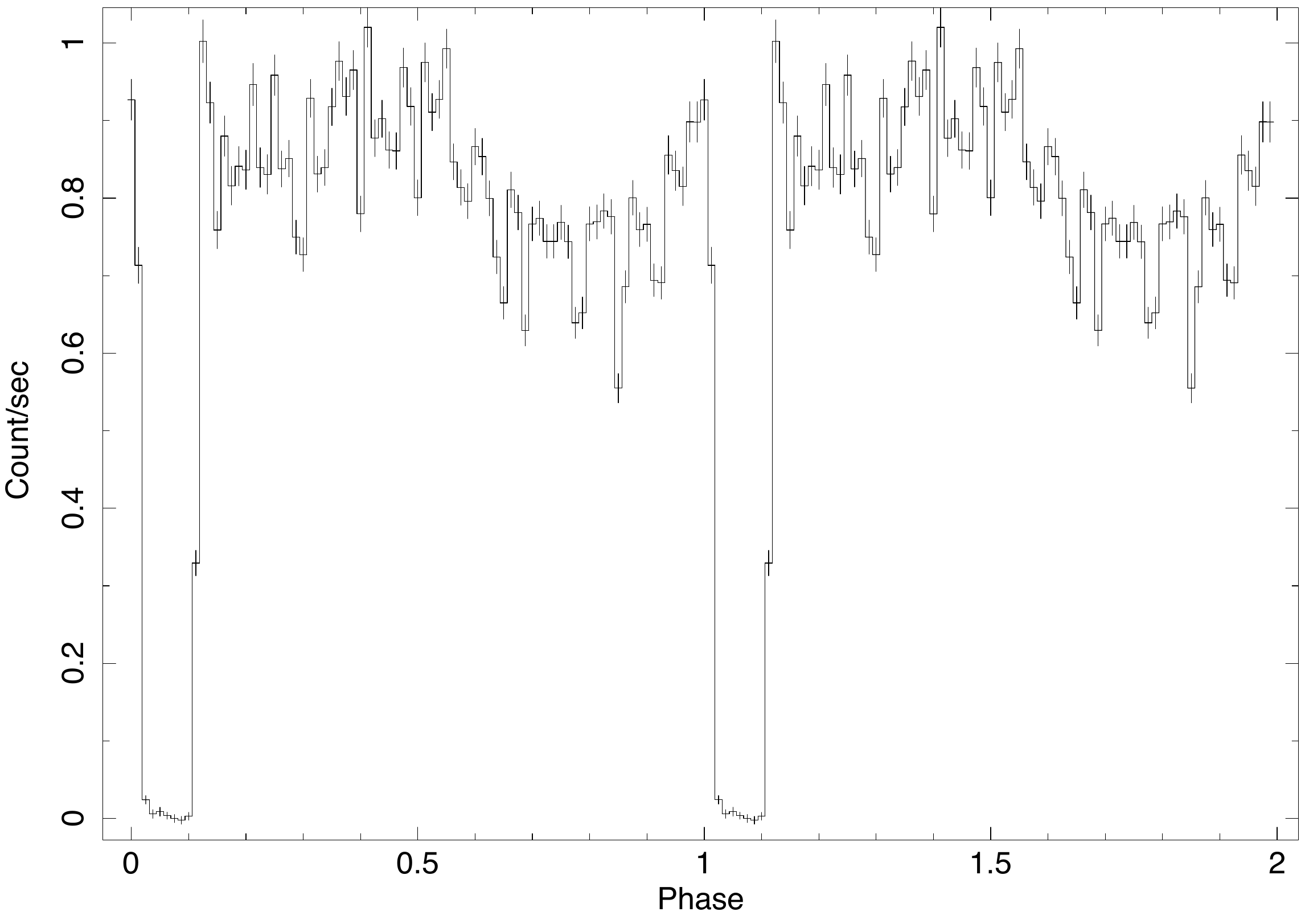}
    \caption{The background subtracted light curve of XY Ari obtained from EPIC pn folded over the orbital period (6.06 h) with 80 phase bins (0.2-10 keV). The binning time is 272.91 s.}
    \label{Fig. 2}
\end{figure}

The background subtracted X-ray light curves of the source were extracted using all the channels of EPIC pn and MOS(1,2) instruments with a time binning of 2 s in the 0.2-10 keV energy band (see Fig.1). The binning time of 100 s is adopted in Fig. 1 for the sake of clarity. The light curve was folded over the orbital period using near-infrared ephemerides given by \cite{Allan96} $HJD=244 7751.8398(1)+0.25269664(6)\times E$ ($E$ is the number of cycle) (see Fig. 2)\footnote{We phase-locked the X-ray light curve to the infrared light curve. The error accumulated from phase locking amounts to 0.0072.}. Phase zero corresponds to mid-eclipse, when the red dwarf lies along the observer's line of sight. As can be seen from Fig.2, there is a slight phase offset between the phase zero of the infrared and X-ray observations (0.067 in phase). Following the method of Hellier, linear fitting with constant terms was applied for the eclipsed and the uneclipsed sections of the light curve. The egress and ingress times were fitted with linear relations with variable slopes. This method yielded a phase of mid-eclipse at 0.067(4). A rough estimate of the egress and ingress times were obtained as $54.58\pm16.11$ s and $34.83\pm13.98$ s, respectively using a binning time of 10 s. The duration of these times are in accordance with the previous results within errors obtained by \cite{Kamata91} (50 s) and \cite{Hellier97} ($25.8^{+1.5}_{-1.8}$ s)\footnote{Only the time of egress could be measured in \cite{Hellier97}.}. The start time of the first data point in the X-ray light curve of EPIC pn and MOS corresponds to the orbital phase 0.255(4) and 0.193(4), respectively. The folded light curve of EPIC pn shows that the X-ray count rate during mid-eclipse time (minimum count of 0.0042 ct/s) is less than 0.5\% of the non-eclipsed light (persistent emission).

\begin{figure}
\centering
	\includegraphics[angle=0, width=0.90\columnwidth]{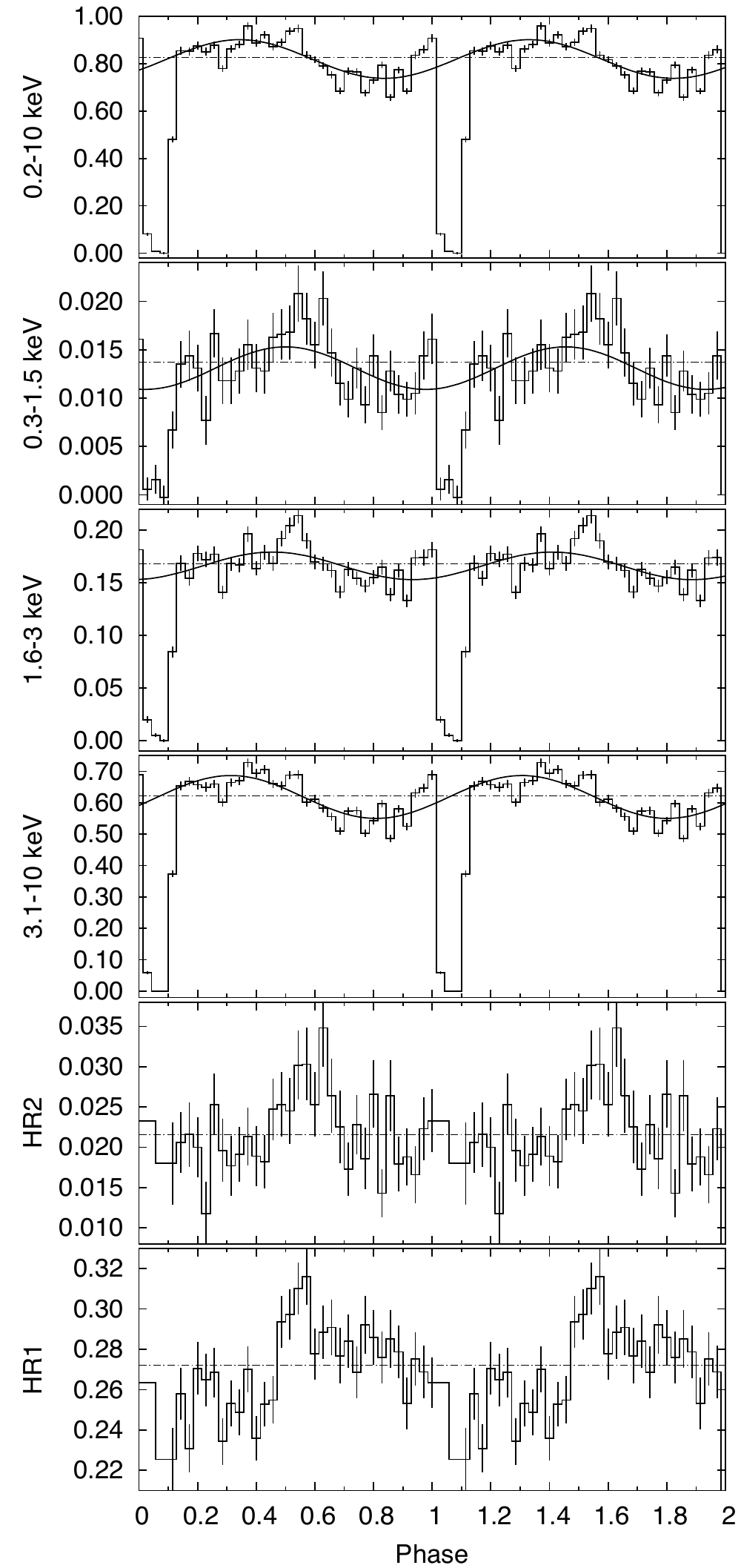}
    \caption{The EPIC pn light curves of XY Ari extracted from different energy ranges (0.2-10 keV (a), 0.3-1.5 keV (b) , 1.6-3.0 keV (c) , 3.1-10 keV (d))  and folded at 6.06 h orbital period of the binary. The solid lines show the sinusoids which are the best fit to each light curve (for details see the text). The average count rates of the broad orbital modulations are shown with the dashed lines. Hardness ratio as a function of the phase derived from the light curves in 0.3-1.5 and 3.1-10 keV (e), 1.6-3.0 and 3.1-10 keV (f). The \textit{y}-axis represents the count rate per second. The binning time is 2 s.}
    \label{Fig. 3}
\end{figure}

Additionally, the light curves for three different energy ranges were extracted in order to search for the energy dependence of the orbital modulation. The selected ranges were 0.3-1.5 keV (panel b), 1.6-3.0 keV (panel c), 3.1-10 keV (panel d) for soft, intermediate and hard energies, respectively. The extracted light curves folded over the orbital period are shown in Figure 3. We fitted a sinusoid to these data (excluding the eclipse itself) with amplitude, mean level and phase of minimum flux as free parameters. The sinusoidal fits overplotted in Fig. 3 for each case. To estimate the energy dependence of the broad orbital modulation and eclipse depth, we defined the percent modulation depth as ((max - min)/ (max + min))$\times100$, where ``max" and ``min" are the pulse maximum and minimum count rates, respectively. The average persistent emission count rate in each light curve, which are overplotted in Fig. 3, were determined as $0.013\pm 0.004$, $0.16\pm 0.01$, $0.62\pm 0.02$ and $0.83\pm 0.02$ cts/s for 0.3-1.5 keV, 1.6-3.0 keV, 3.1-10 keV and 0.2-10 keV energy ranges, respectively. The depth of the broad orbital modulation ($18\pm5$ \%, $8\pm2$ \% and $11\pm2$ \% for soft, intermediate and hard energies) are significant at the $\sim 4\sigma$ level in the hard energy bands ($\geq$ 1.6 keV). The hardness ratios [HR1=((1.6-3.0)/(3.1-10) and HR2=(0.3-1.5)/(1.6-3.0)], that are defined as the ratio soft to hard energies, were derived from light curves for the selected ranges. In Fig.3, both hardness ratios show a softening (increase) at phases between 0.5 and 0.9 and a hardening (decrease) at phases between 0 and 0.5. This behaviour is more prominent in HR1.

\section{Spin Modulation}
We also folded the light curve over the spin (206.298 s) period \citep{Kamata91} using the epoch ($T_0=2450277.40382$ (TDB)) given by \cite{Hellier97}. As proposed by \cite{Hellier97} the epoch corresponds to phase 0.25 where the lower pole is pointing away from the observer and the upper pole is starting to appear. The spin folded light curve of the XY Ari is shown in Figure 4 for EPIC pn (upper panel) and combined MOS (lower panel) in the 0.2-10 keV band. The eclipse times have been excluded from the data to remove the orbital variation with a period of 6.06 h in the flux variation before folding process. The folded light curves of XY Ari show two features. The expected double peaked pulse profile can be seen in both EPIC instruments and is more prominent in the MOS. There is a peak near phase range 0.4-0.5 that arises from the stronger pole. In addition to this, a second peak in the phase range 0.9-1.0 that arises from the weaker pole is detected. Simple sinusoidal fits for MOS data in Fig. 4 show that the spin minima are at phase $0.20 \pm 0.09$ and $0.70 \pm 0.09$. When these are compared with the ones in light curve folded on the spin period between our and Hellier's observations. There is a $\sim$ 0.2 shift in phase between two observations. The accumulated error in phase is 0.0851 that is very similar to our phase error which is calculated by quadratically adding the phase errors due to time bin size of the light curve and the accumulated phase errors in time. As a result, there seems to be some shift in spin phase between different epochs. As can be seen in Fig. 4, the count rate does not drop to zero count rate during the minimum phases, consistent with double pole accretion where there is always some emission visible from one of the poles as one appears and the other disappears depending on the magnetic axis inclination.

The second peak was found to be less prominent in EPIC pn and we examined this as follows. The entire observation was divided into six consecutive data segments after the eclipse times were excluded. Each data set renamed starting with the letter `` a " out to `` f " and folded over the spin period of XY Ari (see Fig. 1). We noticed that there is a distinct difference in pulse profiles and the depth of the second pulse was shallow in some data segments. We decided to sort out the EPIC pn and MOS data into two groups that show similar pulse profiles and these groups were designated as \mbox{`` a+e+f "} (red colored data sets in Fig.1) and `` b+c+d " (blue colored data sets in Fig.1) (see  Fig. 5). As can be seen in the figures, the spin folded pulse profiles of b+c+d group show an apparent double peak profile that has a strong spin maximum at phase 0.4-0.5 and a second maximum at phase range 0.9-1.0. However the pulse profile of group a+e+f represents a single peaked pulse profile, particularly in EPIC pn. The modulation depths of the primary peak (secondary peak) of the pulse profiles of a+e+f groups are $0.153\pm0.024$ ($0.062\pm0.025$) (for pn data), $0.199\pm0.023$ ($0.091\pm0.024$)(for combined MOS data).The modulation depths of the primary peak (secondary peak) of the pulse profiles of b+c+d groups are $0.173\pm0.028$ ($0.097\pm0.029$) (for pn data), $0.251\pm0.028$ ($0.154\pm0.029$)(for combined MOS data). The modulation depths of the second peak are small but just significant at the $2.5 \sigma$ level for PN and significant above the $3 \sigma$ level for MOS data. It is noticeable that there is a difference in emission or a modulation depth between two groups. This means that the relative peak heights of the spin profiles, corresponding to different time intervals, vary during the entire observation which may be a result of the very high inclination of XY Ari \citep[$80\degr < i < 84\degr$,][]{Hellier97}. If we assume that both poles are fed roughly equally, we expect to see two bright pulses during quiescence. In XY Ari, for orbital inclinations of $i=80^{\circ}$ to $84^{\circ}$, the inner disk will obscure the lower pole of the white dwarf if its inner radius moves to below 5 or 9 WD radii, respectively. \cite{HelMukBea97} thus proposed that the lower pole of XY Ari would be obscured during outburst, when the inner edge of the accretion disk pushes closer to the WD (due to the gas pressure increasing, allowing the accretion disk to push the magnetosphere in).  This would then explain how XY Ari's pulsations would change to single-peaked during the outburst, vs. double-peaked at other times.  Alternatively, an increase in the disk scale height during outburst could accomplish the same effect. Another explanation could be that the relative peak heights of the spin profiles, corresponding to different time intervals, vary during the entire observation which may be a result of small accretion rate differences or variable scattering from the weaker pole.

To see how the profiles vary with energy, the light curve of XY Ari was folded over the spin period in different energy bands of 0.3-1.5, 1.6-3.0, 3.1-10 keV. In order to maximize the signal-to-noise ratio the combined EPIC pn and MOS data were used. We plotted the normalized spin pulse profile figures in Fig. 6. The pulse profiles are double-peaked in each energy band. Generally, all the spin pulse profiles have a primary peak within errors at phase $0.5\pm 0.1$ and a secondary peak at $1.0\pm 0.1$. The double-peaked shape is more prominent at higher energies (1.6-3.0 and 3.1-10 keV), while the second peak is weaker at low energies (0.3-1.5 keV). 

We calculated the hardness ratios as a function of the spin phase (Fig. 6). HR1 and HR2, which are defined in Sec. [3], show a prominent softening (increase) at spin maximum of the primary peak and a hardening (decrease) at spin minimum. The hardening is getting greater when the strong pole is disappearing and the weaker pole is appearing, which is prominent in HR1. The hardening at spin minimum is typically observed in IPs and is generally produced by  the larger photoelectric absorption when viewing along the accretion curtain \citep{Rosen88}. 

We noticed that the unmodulated flux level at low energies (0.3-1.5 and 1.6-3.0 keV) has a similar count rate within errors, while there is a significant level difference for the two pulse minima at phase 0.2 and 0.6-0.8 in the high energy range (see Fig. 6). The percent modulation depth which is defined in Sec. [3] can be used to investigate the energy dependence of the pulse amplitude. The modulation depths of the strong peaks are determined as $35(5) \%, 16(1)\%, 11(1) \%$ in 0.3-1.5, 1.6-3.0, 3.1-10 keV range, respectively. This indicates that the modulation depth of this peak decreases with energy. We checked the same relation for the weaker peak. The modulation depths are calculated as $13(6)\%, 6(1)\%, 7(1)\%$ for low to high energy ranges. We suggest spin modulation is energy dependent in general for emission from both of the poles.

During the eclipse, the count rate does not drop to zero. In order to examine whether there is a possible link between the excess X-ray flux and pulse period, we consider the observed excess emission can be resulted from the pulses from both poles of WD. Hence the eclipse times were extracted from the \textit{XMM-Newton} observation. To improve the S/N, combined EPIC pn, MOS1 and MOS2 eclipse data were folded over the spin period using the ephemeris of \cite{Hellier97}. A weighted linear fit is applied to the data, the reduced chi-square is calculated 1.1003 which means there is no evidence for variation in the data in eclipse.

\section{The Phase Averaged Spectrum of XY Ari}

The average EPIC pn and MOS(1,2) spectra were grouped to have a minimum of 100 counts per bin in order to increase S/N. We simultaneously fitted the EPIC pn and two MOS spectra with several models in order to derive spectral parameters like temperature in the (post-)shock accretion column and the amount of intrinsic absorption. The background subtracted X-ray spectra have been analyzed by means of multi-component models in {\scriptsize XSPEC} v12.8 in the 0.3-10 keV energy range. The EPIC pn and MOS(1,2) spectra are shown in Fig. 7. The fitted EPIC spectra show the typical components of IPs; a continuum plasma emission component with superimposed emission lines. The iron line complex around 6.7 keV, which is the strongest emission feature in the X-ray spectra of magnetic CVs, can be clearly detected in EPIC pn and MOS spectra.

We modelled the time-averaged spectrum with single/double/triple thermal plasma emission models ({\scriptsize MEKAL} and {\scriptsize APEC/VAPEC} models in {\scriptsize XSPEC}). These models assume collisional equilibrium \citep*{Mewe85}. A continuous distribution of temperatures can also be modelled using the {\scriptsize CEVMKL} \citep{Done97}. We added a Gaussian line at 6.4 keV to account for the reflection component. An absorber model for the interstellar medium ({\scriptsize TBabs}, Tuebingen-Boulder ISM, \cite{Wilms00}) and a partially covering photoelectric absorber ({\scriptsize PCFABS}) for intrinsic absorption were assumed.

\begin{figure}
\centering
	\includegraphics[angle=0, width=0.85\columnwidth]{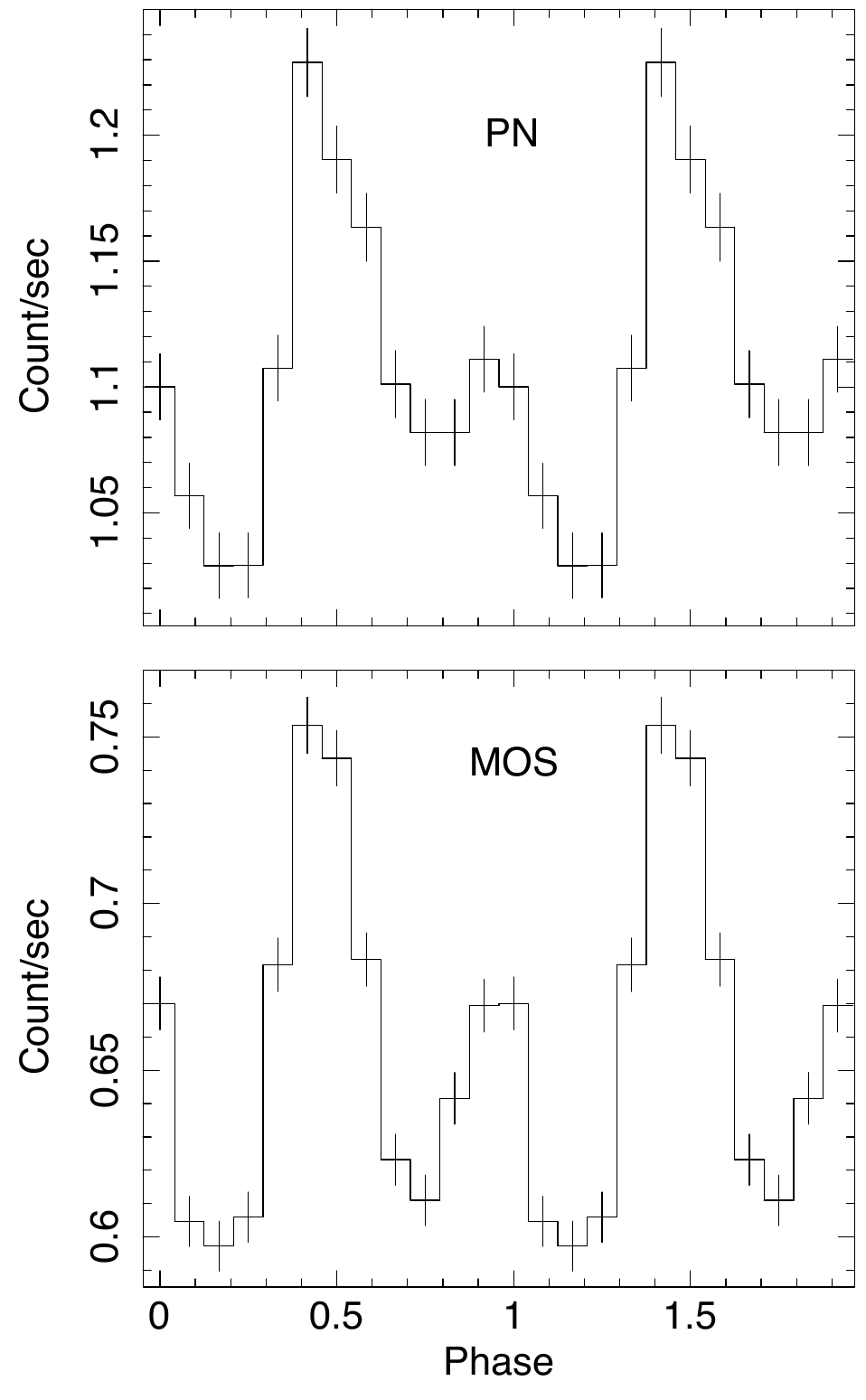}
    \caption{The light curve of EPIC pn (top) and MOS (bottom) folded over the spin period of 206 s using the ephemeris $T=2450277.40382+E\times0^d.00238771(6)$. In both panels the energy range is 0.2-10 keV.}
    \label{Fig.4}
\end{figure}

\subsection{Line of Sight Absorption}

Since XY Ari lies behind the molecular cloud known as MBM 12, the optical counterpart couldn't be observed. It has long been thought that MBM 12 was the nearest molecular cloud with a distance of 65 pc \citep{Hobbs86}. Recent studies suggest larger distance values of 360$\pm$30 pc \citep{Andersson02}, 275$\pm$65 pc \citep{Luhman01}, and 325 pc \citep{Strai02}. In addition to this, using infrared observations, XY Ari has been found to be at a distance of 270$\pm$100 pc with a high visual extinction and thus must be located behind the molecular cloud MBM 12 \citep{Littlefair01}. The photoelectric absorption of molecular cloud was calculated by \cite{Smith05} in the range $(6-8.4)\times 10^{21} \,\mathrm{cm^{-2}}$. \cite{Heit10} obtained a bolometer map of the densest part of the cloud at a wavelength of 1.2 mm with the \textit{IRAM} 30 m telescope. The molecular hydrogen column density was derived to be $N_\mathrm{H} = 1.3\times 10^{22} \,\mathrm{cm^{-2}}$. The \textit{XMM-Newton} observation of MBM 12 was analyzed by \cite{Kout11} and the obtained $N_\mathrm{H}$ value was given as $0.7(\pm2)\times 10^{22} \,\mathrm{cm^{-2}}$. 

In this study we noticed that when we used a single absorber model for interstellar extinction ({\scriptsize TBabs}) and left it to vary freely, the best-fit value of $N_\mathrm{H}$ was in range $(3.5-4)\times 10^{22} \,\mathrm{cm^{-2}}$ (at 90 \% confidence level). This column density is higher than the interstellar value indicating a characteristic intrinsic origin as found in many IPs. Therefore, a second intrinsic absorption component that fully ({\scriptsize TBabs}) or partially ({\scriptsize PCFABS}) covers the X-ray emitting region needs to be included in the fitting process.

\begin{figure}
\centering
	\includegraphics[angle=0, width=\columnwidth]{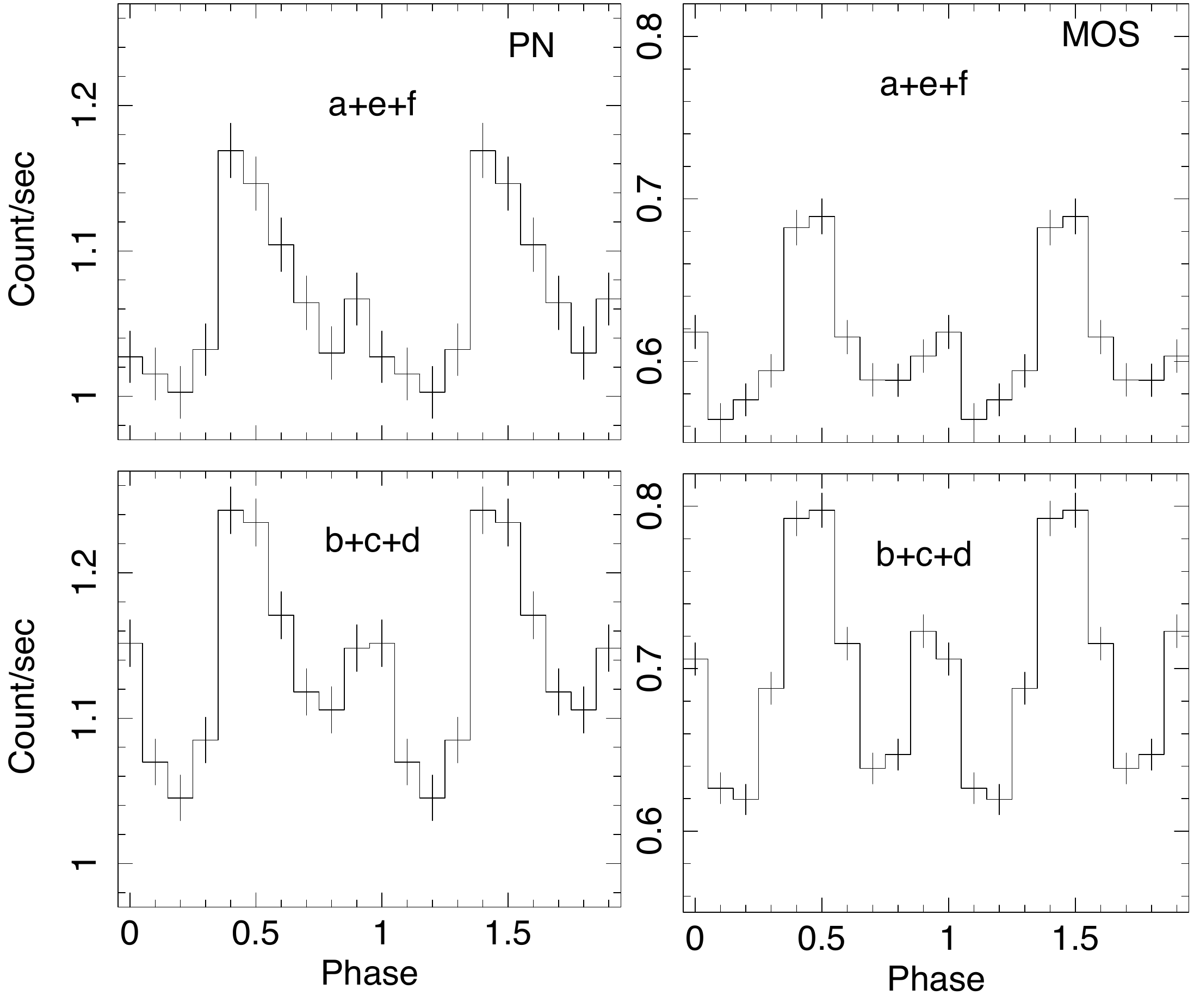}
    \caption{The light curves of the grouped data of EPIC pn (left) and combined MOS (right) folded over the spin period in 0.2-10 keV. The ephemerides are same as in Fig. 4. For details see Sec. [4].}
    \label{Fig. 5}
\end{figure}

\begin{figure}
\centering
	\includegraphics[angle=0, width=0.90\columnwidth]{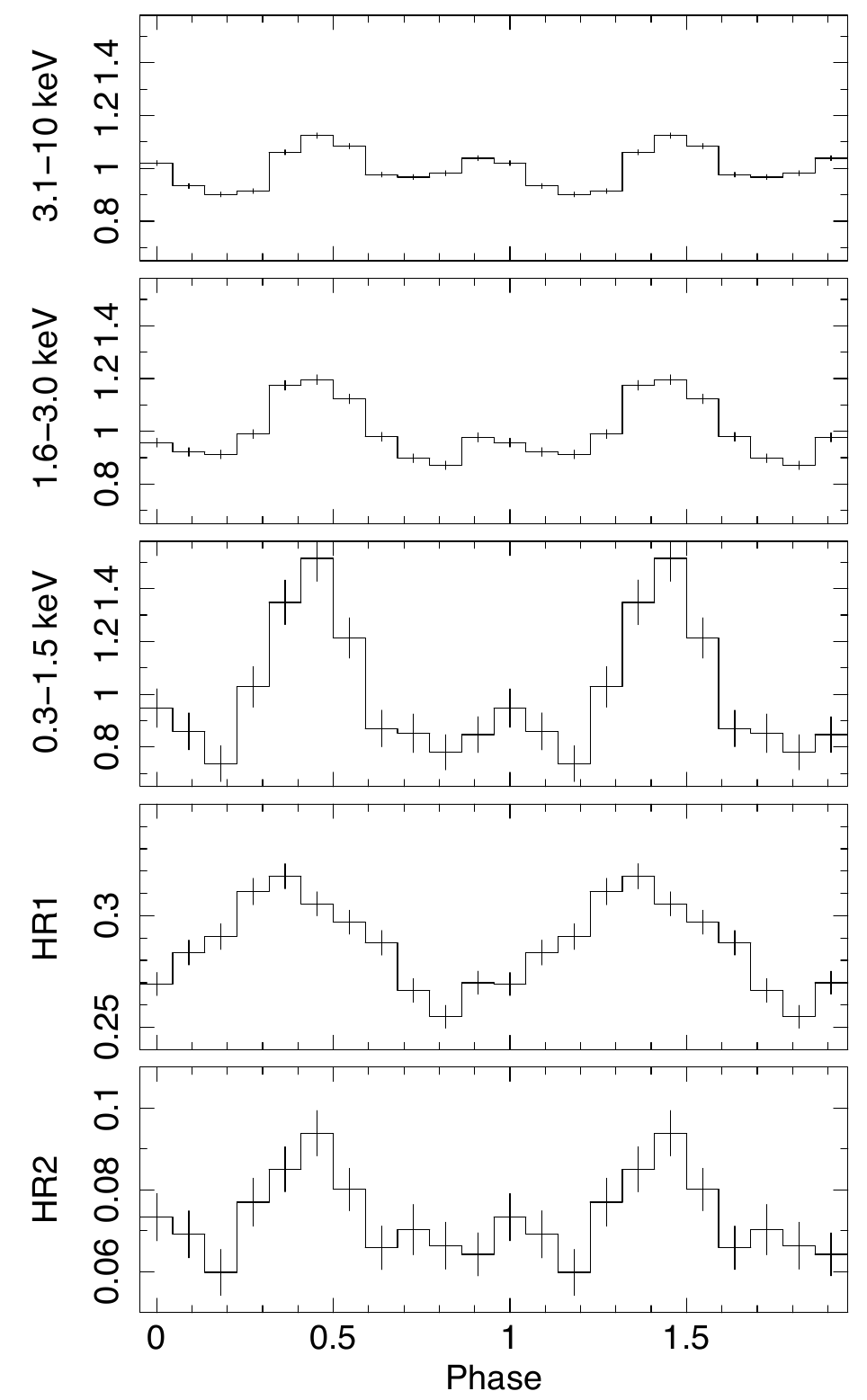}
    \caption{The pulse profile and hardness ratio obtained from EPIC data folded over 206 s spin period for different energy ranges: 0.3-1.5, 1.6-3.0, 3.1-10 keV from top to bottom. The profiles are background subtracted and normalized to the average count rate. The hardness ratios are derived from pulse profiles in two neigbouring energy bands (1.6-3.0 and 3.1-10 keV for HR1,  0.3-1.5 and 1.6-3.0 keV for HR2). The ephemerides are same as in Fig. 4.}
    \label{Fig. 6}
\end{figure}

\subsection{Analysis with CEVMKL Model}

As we have stated at the beginning of Sec. [5.0], we have attempted fits to the entire EPIC data simultaneously with plasma emission models. A single temperature model fit yielded $\chi^{2}_{\nu}$ value of 1.24 and double MEKAL or APEC model fits yielded fits with unconstrained high temperatures for the plasma emission with $\chi^{2}_{\nu}=1.12$. Therefore, we assumed a reasonable estimate for the physical description of the shock-heated plasma emission by assuming a model of multi temperature plasma flow with variable elemental abundances. We used the optically thin collisional equilibrium plasma model CEVMKL in XSPEC to represent the temperature distribution in the post-shock plasma. The CEVMKL model, which is built from the MEKAL model, introduces the power-law type differential emission, proportional to $(T / T_{max})^\mathrm{\alpha}$. The abundances for the 13 most common elements from C to Ni can be set separately in the model. The EPIC pn and MOS spectra are fitted simultaneously (see Table 1). A constant factor is used to account for a normalization uncertainty between the EPIC pn and MOS instruments. It is fixed at unity for the EPIC MOS but is left free for the EPIC pn. When the alpha parameter that is the power law index of the temperature distribution in the plasma flow in the shock-zone was allowed to vary freely, the fit did not constrain the temperature, thus we kept it fixed at 1.0 implying a isobaric cooling flow model as was found in most IPs. When the abundances of all elements were allowed to vary together in the CEVMKL fit, the abundances of some elements were found super solar and the errors of abundances were too high. So the elemental abundance ratios in CEVMKL model are fixed at solar values and only the Fe abundance is left as a free parameter. 

Initially the spectrum was fitted with the $cons \times tbab \times (CEVMKL)$ composite model but the model did not provide a good fit to the data ($\chi^{2}_{\nu} = 1.77/1286$). Adding a partial covering absorber (pcfabs) significantly improve the fit ($\chi^{2}_{\nu} = 1.49/1284$). The CEVMKL model accounts for the thermal lines, i.e. the helium-like iron line at 6.7 keV and the hydrogen-like iron line at 6.97 keV but not for the fluorescent K$\mathrm{\alpha}$ line of neutral iron at 6.4 keV line which originates by reflecting cool material. In addition to this, $kT_{max}$ typically favoured high values (>80 keV). The inclusion of a gaussian line at 6.4 keV significantly improve the fit, when the abundance was kept at solar value ($\chi^{2}_{\nu}  = 1.32/1283$). This model however still resulted in fits with a high plasma temperature of 73.9 keV. A second partial absorber was added to the model motivated by \cite{Yuasa10} and \cite{Ezuka99} that performed the spectral fitting of XY Ari including a partial covering absorption with two or three column densities. The F-test shows that the addition of component to the model improves the fit significantly, with a chance probability of this occurring of $2.7 \times 10^{-6}$ ($\chi^{2}_{\nu} = 1.17/1279$).

Finally we fitted a model consisting of a photoelectric absorber and two partial covering photoelectric absorbers ($tbabs \times 2 pcfabs \times (\mathrm{CEVMKL+GA})$) to the spectrum. The best-fit solution gave a $\chi^{2}_{\nu}$ of 1.17 (1278 d.o.f) and plasma temperature of $28.0^{+3.1}_{-2.9}$ keV with an Fe abundance of $0.37^{+0.06}_{-0.05}$. For the partial covering absorbers we find, $N_{\mathrm{H1}} = 6.2^{+1.0}_{-0.9} \times 10^{22} \,\mathrm{cm^{-2}}$ and $N_{\mathrm{H2}} = 105.3^{+35.4}_{-30.4} \times 10^{22} \,\mathrm{cm^{-2}}$ with covering fraction of 0.53 and 0.41. The intrinsic absorption was determined as $N_{\mathrm{H}} = 3.5^{+0.1}_{-0.2} \times 10^{22} \,\mathrm{cm^{-2}}$. All the errors are at 90 \% confidence level. We calculate an unabsorbed flux as $4.7^{+1.5}_{-0.8} \times 10^{-11} \,\mathrm{erg \,cm^{-2} s^{-1}}$ in 0.3-10 keV range. The equivalent width of the line at 6.4 keV is 51 eV for EPIC pn and MOS. The spectral parameters from the fit result are displayed in Table 1. Figure 7 shows the simultaneously fitted EPIC spectra with the best fitting model described in this paragraph. The spectra around the iron line complex are given in Figure 8. 

The X-ray spectra of XY Ari obtained with \textit{ASCA} fitted by a single-temperature thermal bremsstrahlung with a fixed temperature of 17 keV shows three partial covering absorbers in range of $(3.7-70)\times 10^{22} \,\mathrm{cm^{-2}}$ by \cite{Ezuka99}. This is also consistent with our findings. We find that the model with three partial covering absorbers did not improve the spectral fit significantly with F-test false probabilities of 0.62. In this model, the $N_{\mathrm{H}}$ values were $108.9 \times 10^{22}\, \mathrm{cm^{-2}}$,  $6.6\times 10^{22} \, \mathrm{cm^{-2}}$ and $1.54 \times 10^{22} \, \mathrm{cm^{-2}}$ with covering fraction of 0.48, 0.42 and 0.95, respectively. The parameters of the partial absorbers in our fits are comparable to the values obtained for many IPs.

In addition, our composite model fit indicates that there is some soft excess in the EPIC pn spectrum between 0.6-1.5 keV. Such excess are generally associated with a blackbody component originating from reprocessing of the hard X-rays in the WD atmosphere. It has been known that a distinct black body component in softer energies has been detected in some IPs \citep[e.g.][]{Frank02,deMar04,Bernar12}. The upper limit for temperature was derived $\sim100$ eV in the list of analyzed spectrum of IPs with bbody component by \cite{Eva07} and \cite{Anzo12}. To account for this excess, we added a single-temperature blackbody component to our model with this temperature. The model with a blackbody does not provide a good fit which means the presence of this component was unnecessary in the spectral fit.

However, rather than assuming a potential blackbody emission from the WD surface, it is possible that the scattered X-rays from dust in the molecular cloud of MBM12 could be responsible \textbf{for} the soft excess below 1.5 keV. X-ray photons passing through the interstellar medium are subject to scattering by dust grains, which affects soft X-rays the most strongly (e.g. \cite{Drain03}). A fraction of soft X-rays initially traveling away from us can be scattered in our direction, arriving even when our direct view of the source is blocked (e.g. \cite{Xu86}). \cite{Tan04} considered arcsecond scale X-ray scattering halos from Galactic Center sources with the Chandra observations and showed the halo profile changed with the spatial distribution of the dust that is close to the source. \cite{Jin17} investigated the influence of the dust scattering from clouds near the Galactic Centre on the observed spectra of AX J1745.6-2901. They found that the observed spectra of the source can strongly depend on the source extraction region, which is due to the partial inclusion of the dust scattering halo.

\begin{figure*}
	\includegraphics[angle=0,width=0.85\textwidth]{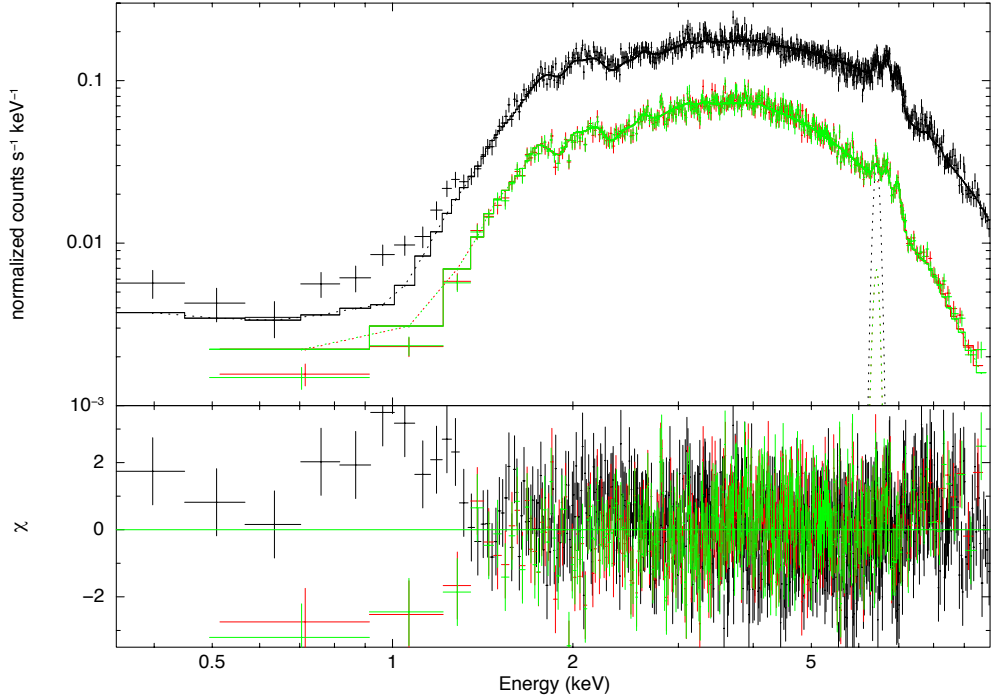}
    \caption{ The phase average EPIC spectra of XY Ari with the best fitting composite spectral model (tbabs*2pcfabs*(CEVMKL+Gaussian)) in the 0.3-10 keV energy band. The crosses show the data with errors, solid lines show the composite model for EPIC pn (black) and MOS (green and red), the dotted lines show the individual models. The bottom panel shows the residuals.}
    \label{Fig. 7}
\end{figure*}

\begin{figure}
	\includegraphics[width=\columnwidth]{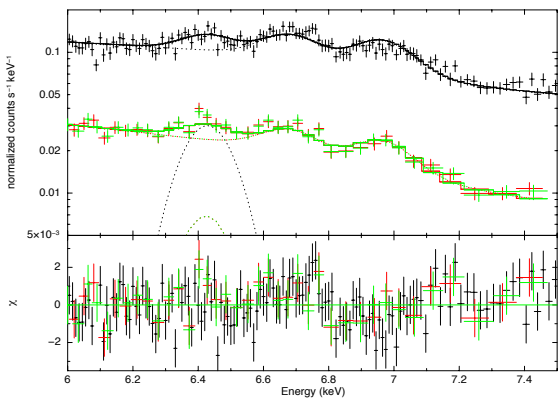}
    \caption{The region from 6 to 7.5 keV including the Fe line complex in the EPIC spectra is shown. The bottom panel shows the residuals. }
    \label{Fig. 8}
\end{figure}

\section{The Spin Phase-Resolved Spectra}

We also performed spin-phase resolved spectroscopy using the ephemerides given in Sec. [4]. In order to investigate variations with the spin cycle, we excluded the data from the orbital dips. First, we created spectra of the phases 0-0.1, 0.1-0.2, 0.2-0.3, .., 0.9-1.0 for each using data from both the EPIC pn, MOS1 and MOS2 instruments. The calculated spectra were grouped such that each energy bin contained minimum of 40 counts and data points below 0.3 keV and above 10 keV were omitted due to bad statistics. 

Initially, we fitted simultaneously all the spectra of the three EPIC cameras in each phase (30 spectra) using a simple model ($tbabs \times (CEVMKL+Ga)$) tying the CEVMKL temperature, Fe abundance, Gaussian line energy and sigma parameters. These parameterse were fixed to the value found for the average spectrum. The fit yielded $\chi^{2}_{\nu} = 1.67/3514$. The inclusion of a partial covering absorber significantly improve the fit with $\chi^{2}=1.36$ at 3513 d.o.f.. Adding a second partial covering absorber to the model yielded $\chi^{2}_{\nu} = 1.34/3511$. The low value of the F-test probability ($9 \times 10^{-12}$) favours this additional model component. However it is important to  realize that partial covering absorbers may be partial in space (e.g. part of the emitting region is covered by absorber) or time (absorbers pass over the region during part of the time in which spectra are accumulated). In the latter case, the obtained spectrum may not be distinguished from spatial partial covering. Thus we adopt a single partial covering absorber to fit the data as the simplest convenient representation for physical interpretation. We fitted simultaneously the spectra of the three EPIC cameras using $tbabs \times pcfabs(CEVMKL+Ga)$ tying all the parameters at their best-fitting values except for the photoelectric absorber (TBABS) NH and partial covering absorber (pcfabs) NH values and evaluated the physical parameters quantitatively in each phase. Figure 9 shows the selected two parameters versus the spin phase together with the folded light curve for the 0.3-10.0 keV energy band of the EPIC pn data (see also Table 2). 

In general, the $\textit{N}_{H}$ parameter for the intrinsic absorption and partial covering absorber show anticorrelation with the flux. They yield higher values at spin minimum phases (phases between 0 and 0.3; between 0.6 and 0.9) and lower values at the maximum phases (phases between 0.3 and 0.6; 0.9-1.0). This indicate that source emission is being absorbed at particular phases (0.1-0.3 and 0.7-0.9). The decrease in the $\textit{N}_{H}$ parameter for the partial covering absorber ($\sim 30 \%$) seems much greater than the $\textit{N}_{H}$ parameter for intrinsic absorption ($\sim 8 \%$) which may indicate that the contribution of an accretion column to total absorption seems important. Additionally when the absorption column for the partial covering absorber seems more dominant at phases 0.1-0.3 than the phases 0.6-0.8, it is just the opposite for the simple photoelectric absorber (TBABs). It seems that the absorption column for TBABs is efficient on broad phase interval (phases between 0.7-1.3) peaking at phase 0.7. While the $\textit{N}_{H}$ parameter for the partial covering absorber yields lower values at tha maximum phases (0.9-1.0), on the contrary, the simple photoelectric absorber (TBABs) yields higher values in this phase range. This may indicate that the emission coming from the lower pole is being absorbed by the simple photoelectric absorber at this particular phases. As a result, the total effect of the simple absorber and partial covering absorption on the line of sight to lower pole seems leading far more absorption which can explain the modulation of the two pulses discussed in Sec. [4].

\section{Discussion}

Mukai's list of confirmed IPs\footnote{http://asd.gsfc.nasa.gov/Koji.Mukai/iphome/iphome.html} contains 47 systems where five IPs (EX Hya, V598 Peg, TV Col, BG CMi, FO Aqr, CC Scl) show partial or grazing eclipses while two IPs (XY Ari, DQ Her) have been identified as deeply eclipsing systems\footnote{V597 Pup \citep{Warner09} and IPHAS J0627 \citep{Aun12} have also been reported as deeply eclipsed intermediate polars, but these systems are marked as possible or probable IPs in the catalog of IPs and IP candidates. Swift J201424.9+152930 \citep{Espo15} and CXOGBS J174954.5-294335 \citep{John17} are reported as newly confirmed deeply eclipsing intermediate polars.} suitable to provide detailed information about the accretion column structure of an IP. But XY Ari is the X-ray brightest deeply eclipsing IP and it exhibits almost total eclipses. It is a well studied system with a spin period of 206.298 s and an orbital period of 6.06 hrs. 

The results from the folded X-ray light curve of XY Ari over the orbital period show clearly the deep eclipses with the X-ray flux virtually disappearing for $\sim 10 \%$ of the orbital period. The mid-eclipse is found at phase $0.067\pm0.004$ by fitting the folded light curve. The difference in the mid-eclipse time of the X-rays and the IR may arise from the fact that it is a very sharp and deep variation which also depends on the geometry of the accretion region on the WD and the accretion curtains. Another possible origins behind such variations could be related with the mode of accretion such as a variable mass accretion rate, changes in the activity of the secondary or changes in the structure of the accretion disc. For instance, possibly the orbital period changes slightly due to changes in the mass accretion rate, which would then lead to phase shifts comparing the X-ray observation to the IR ephemeris from previous observations. Additionally, the magnetic activity can cause the late-type companion star to show cool spots on its surface and the asymmetries in the infrared light curve may cause the time of minima to be slightly shifted in time.

\begin{figure}
\centering
	\includegraphics[angle=0, width=0.95\columnwidth]{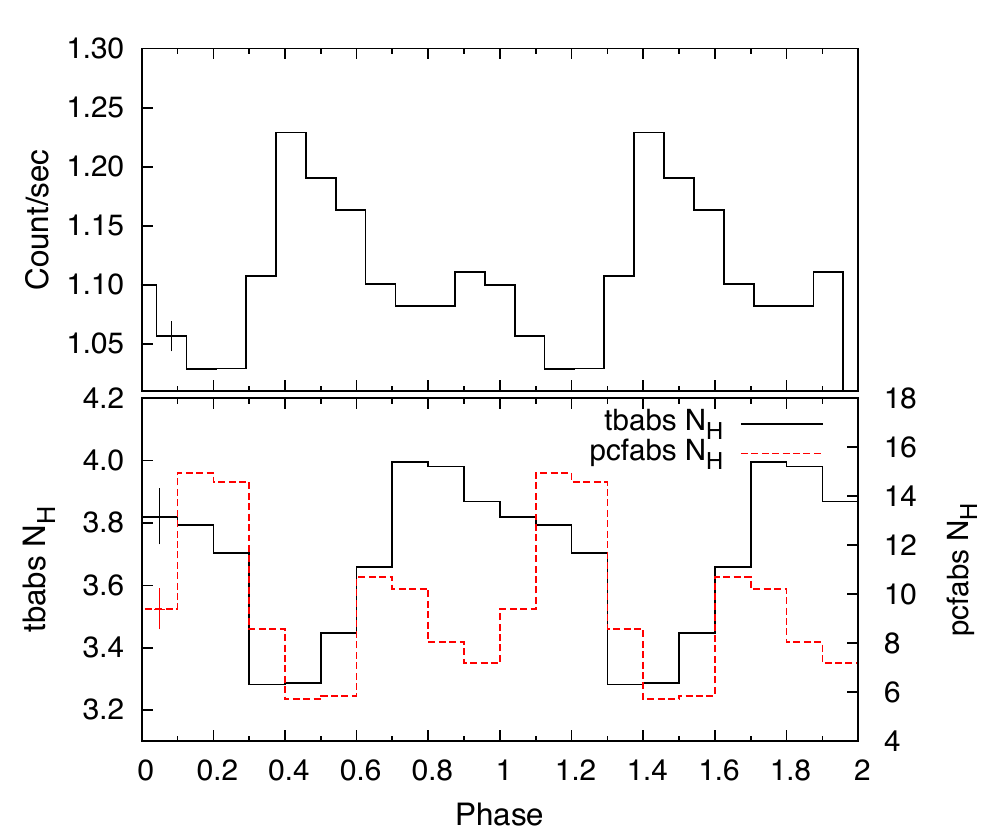}
    \caption{The light curve of EPIC pn folded over the spin period and plots of two spectral parameters derived from the spin phase-resolved spectroscopy over the same spin phase range. $N_{\mathrm{H}}$ values are in units of $\times 10^{22} \,\mathrm{cm^{-2}}$. The error bars correspond to $2\sigma$ confidence level. The ephemerides are same as in Fig. 4.}
    \label{Fig. 9}
\end{figure}

Energy dependent-features modulated over orbital period are common in IPs \citep{Hellier93,Parker05}. The cause of the orbital modulation is commonly believed to be local absorption of X-rays in an extended structure at the accretion stream impact zone in the outer edge of the disk \citep[e.g.][]{Parker05,Pekon11,Pekon12}. Nevertheless as seen from the orbital light curves, one can notice cycle-to-cycle variations in the out-of-eclipse data which could affect the results. We can safely ascribe to the photoelectric absorption in a local structure near the outer edge of the accretion disc across the observer's line of sight.

\begingroup
\setlength{\tabcolsep}{10pt} 
\renewcommand{\arraystretch}{1.5}

\begin{table*}
 \caption{The spectral parameters of XY Ari obtained from a fit with the composite model ($tbabs \times 2 pcfabs \times (\mathrm{CEVMKL+GA})$) in the 0.3-10.0 keV range using the EPIC pn, MOS1 and MOS2 data. The given errors correspond to $2\sigma$ confidence level for a single parameter. Flux has been determined via {\scriptsize CFLUX} within {\scriptsize XSPEC}. The constant parameter represents the normalization uncertainty between the EPIC pn and MOS instruments.}

 \label{tab:Table 1}

  \begin{tabular}{llll}
  \hline
  Model & Parameter & Unit & Value (Error) \\
  \hline\hline
  tbabs & $N_{\mathrm{H}}$ & $\times 10^{22} \,\mathrm{cm^{-2}}$ & $3.5^{+0.1}_{-0.2}$ \\
  pcfabs &$N_{\mathrm{H}}$ & $\times 10^{22} \,\mathrm{cm^{-2}}$ & $6.2^{+1.0}_{-0.9}$ \\
               & Cvr.Frac. &  & $0.53^{+0.05}_{-0.04}$ \\
  pcfabs &$N_{\mathrm{H}}$ & $\times 10^{22} \,\mathrm{cm^{-2}}$ & $105.3^{+35.4}_{-30.4}$ \\
               & Cvr.Frac. &  & $0.41^{+0.14}_{-0.13}$ \\
  CEVMKL& $kT_{max}$ & keV & $28.0^{+3.1}_{-2.9}$  \\
               & Fe abun. & & $0.37^{+0.06}_{-0.05}$ \\
               & Norm. ($\times 10^{-2}$) & & $5.03^{+0.02}_{-0.01}$ \\
  Gaussian & Line E & keV & $6.43^{+0.01}_{-0.02}$ \\
             & $\sigma$ ($\times 10^{-3}$) & keV & $6.28^{+0.04}_{-0.01}$  \\
             & Norm. ($\times 10^{-5}$) & & $1.81^{+0.06}_{-0.36}$ \\
  Constant & EPIC pn &  & $0.96^{+0.01}_{-0.01}$ \\ 
   & EPIC MOS & & 1.0 (fixed)    \\       
 $Flux_{(0.3-10 \,\mathrm{keV})}$ & & $\times 10^{-11} \,\mathrm{erg \,cm^{-2} s^{-1}}$ & $4.7^{+1.2}_{-0.7}$\\
 $\chi^{2}_{\nu} (\mathrm{d.o.f}) $ & & & 1.17 (1278) \\   
  \hline

  \end{tabular}
  
\end{table*}
\endgroup

It is reported that XY Ari displays a broad, sinusoidal X-ray orbital modulation which is superimposed on the deep X-ray eclipse \citep{Kamata91,Salinas04,Norton07} and it varies in phase and significance. \cite{Norton07} noted this modulation was prominent in the various observations (\textit{Ginga, ASCA, Chandra} and \textit{XMM-Newton} between 1989 and 2000) then it disappeared entirely in 2001 (\textit{XMM-Newton}) and it couldn't be detected with any significance in 2005-2006 (\textit{RXTE}). We investigated the broad orbital modulation to compare with previous observations. The hard X-ray (greater than 1.6 keV) light curves showed energy-dependent variations at above $3 \sigma$ significance. Unlike the results of the last observations reported by \cite{Norton07}, XY Ari seemed to show the broad orbital modulation again; 10 years. The causes of the broad orbital modulation in XY Ari are investigated in the comprehensive study of \cite{Norton07}. In their study, it was suggested a precessing, tilted accretion disc may responsible for the changes in the observed modulation. Such a tilted disc will raise bulge at its outer edge in orbital phase. The lifted material moves into and out of our line of sight. Thus a varying presence or depth of X-ray orbital modulation would naturally arise. A tilted precessing disc is observed in the IP TV Col which has a 4 d disc precession period \citep{Barret88}. 

Spin variations in the X-ray light curve of IPs are often the result of absorption as accretion curtains of material wander across the observer's line of sight \citep[e.g.][]{Hellier91,Kim95,Evans04}. We find that the spin profile is double-peaked in the 0.2-10.0 keV range which is similar to that reported by the previous studies \citep{Kamata93,Hellier97}. The asymmetries between the poles which were reported using the \textit{RXTE} observation of the source by \cite{Hellier97} can be clearly seen in the EPIC pn observation.

\begingroup
\setlength{\tabcolsep}{10pt} 
\renewcommand{\arraystretch}{1.5}

\begin{table*}
 \caption{The two spectral parameters of XY Ari derived at each spin phase of 0.1 in the 0.3-10 keV range. A fit with the composite model ($tbabs \times pcfabs \times (\mathrm{CEVMKL+GA})$) is used for all the spectra. The given errors correspond to $2\sigma$ confidence level for a single parameter. Gaussian line centre, sigma, CEVMKL temperature and the iron abundance parameter were fixed at values which were derived from the average spectrum. Normalization parameters of two models were fixed their best fitting value.}

 \label{tab:Table 2}

  \begin{tabular}{lllllll}
  \hline
  Model & Parameter & 0.1 & 0.2 & 0.3 & 0.4 & 0.5  \\
  \hline\hline
  tbabs & $N_{\mathrm{H}}$ ($\times 10^{22}$) & $3.79^{+0.08}_{-0.08}$ & $3.70^{+0.08}_{-0.08}$ & $3.28^{+0.08}_{-0.07}$ & $3.29^{+0.09}_{-0.08}$ & $3.45^{+0.09}_{-0.08}$ \\
  pcfabs &$N_{\mathrm{H}}$ ($\times 10^{22}$) & $14.95^{+1.38}_{-1.26}$ & $14.59^{+1.35}_{-1.24}$ & $8.57^{+0.79}_{-0.74}$ & $5.72^{+0.58}_{-0.57}$ & $5.86^{+0.62}_{-0.60}$\\              
 $\chi^{2}_{\nu} (\mathrm{d.o.f}) $  & & 1.02 (330) & 1.20 (336)& 1.07 (384) & 1.18 (416) & 1.02 (403) \\   
  \hline
Model & Parameter & 0.6 & 0.7 & 0.8 & 0.9 & 1.0  \\
  \hline\hline
  tbabs & $N_{\mathrm{H}}$ ($\times 10^{22}$) & $3.66^{+0.08}_{-0.08}$ & $3.99^{+0.10}_{-0.09}$ & $3.98^{+0.10}_{-0.10}$ & $3.87^{+0.10}_{-0.09}$ & $3.82^{+0.09}_{-0.09}$ \\
  pcfabs &$N_{\mathrm{H}}$ ($\times 10^{22}$) & $10.71^{+0.99}_{-0.92}$ & $10.22^{+0.95}_{-0.89}$ & $8.04^{+0.83}_{-0.79}$ & $7.19^{+0.74}_{-0.71}$ & $9.39^{+0.82}_{-0.87}$ \\                                       
 $\chi^{2}_{\nu} (\mathrm{d.o.f}) $ & & 1.16 (356) & 1.08 (347) & 0.97 (356) & 1.15 (371) & 1.07 (357)\\     
  \hline

  \end{tabular}
  
\end{table*}
\endgroup

Comparing the spin profiles of XY Ari with \cite{Hellier97} shows that there is a phase shift by $\Delta\phi\sim$ 0.2 between the two observations. As noted by \cite{Salinas04}, a similar phase shift ($\approx$ 0.3) has been previously noted in the \textit{Chandra} observations where the ephemerides of \cite{Hellier97} was, also, used. We have calculated an accumulated error around 0.1 in phase for the spin modulation, which does not account for the 0.2 phase difference we detect in the profiles. 

We find that the count rate does not drop to zero during the pulse minimum which means the appearance of one pole compensates for the disappearance of the other. As proposed by \cite{Evans04} the absorption and electron scattering in a highly ionized post-shock region have an effect on the opacity which can explain the relative variation in brightness from the view of the accretion curtain. The detailed analysis of the entire observation shows that the relative peak heights of the spin profiles vary during the \textit{XMM-Newton} observation which may be a result of small accretion rate differences or most probably partial obscuration of the weaker pole as a consequence of the high inclination.

The observed variation in spin modulation depth with energy is as expected if the modulation is produced by phase-varying photoelectric absorption close to the surface of the white dwarf in the accretion curtain \citep{Norton89,Norton07}. The behaviour of a decreasing modulation depth with increasing energy has been observed in many IPs \citep{Bernar12}. The energy-resolved spin pulse profiles of XY Ari show a decreasing modulation with energy. This behavior is clearly visible for the strong peak at phase $\sim 0.5$. However, we find that this is only evident at low energies ($\leq 1.6$ keV) for the second weaker pole where the variation seems to change with energy. Thus, we suggest that we may be seeing the scattered light from the weaker pole which hides behind the accretion curtain and the disk for most of the time because of the high inclination of XY Ari. The absorption through the line of sight to lower pole is greater than the line of sight to upper pole which may explain the weak energy dependence. The \textit{Ginga} observation of XY Ari in 1989 showed a decreasing modulation depth with increasing energy \citep{Norton07}. Despite the noisy pulse profile, \cite{Salinas04} remarked that there is a sign for a varying hardness ratio with phase in \textit{Chandra} data in 2000. However \cite{Norton07} pointed out that there was no evidence for energy dependence in the \textit{RXTE} observation of XY Ari in 2005/2006. They suggested a precessing disc could be responsible for the energy-independent pulse profiles. 

The average spectrum of the source can be best fit using a composite model ($tbabs \times 2 pcfabs \times (\mathrm{CEVMKL+GA}$)). This yields  a temperature $kT_{max}=28.0^{+3.1}_{-2.9}$ keV in accordance with the plasma temperatures obtained from other IPs \citep{Butters08,Middle12,Bernardini15}. The Fe abundance was obtained as $0.37^{+0.06}_{-0.05}$ times the solar value. In most previous work with different observations, the spectra of XY Ari was fitted by thermal Bremsstrahlung \citep{Kamata93,Ezuka99,Salinas04,Norton07}. The plasma temperatures estimated in these studies are all almost consistent with our study within the errors. \cite{Yuasa10} estimated the masses of nearby white dwarfs in IPs that was based on X-ray spectral analysis using \textit{Suzaku} in 3-50 keV energy band. A spectral emission model of IPs with resolved Fe emission lines was constructed in this study. The mean Fe abundance and the shock temperature of XY Ari was determined as $0.43^{+0.04}_{-0.03}$ times the solar and $61.0^{+26.2}_{-16.2}$ keV, respectively. This temperature is larger than our value but consistent with the Fe abundance we have derived. However due to the lack of coverage above 10 keV with \textit{XMM-Newton}, the inferred maximum temperature is likely not the shock temperature. The shock temperature depends on the mass of the WD and it is found to be massive in XY Ari. \cite{Yuasa10} derive 1.04 $M_{\sun}$ from the broad-band spectrum and \cite{Hellier97} estimated a mass in the range 0.78-1.17 $M_{\sun}$ (or 0.91-1.29 $M_{\sun}$) from the eclipse study. A 1 $M_{\sun}$ WD would imply a shock temperature of the order of $\sim$ 60 keV.

The best-fit model consisting of a plasma emission component that is CEVMKL and two different partially covering absorber components, have $N_{\mathrm{H}}$ values of $105.3^{+35.4}_{-30.4} \times 10^{22}\, \mathrm{cm^{-2}}$ and $6.2^{+1.0}_{-0.9} \times 10^{22} \, \mathrm{cm^{-2}}$ with covering fraction of $0.41^{+0.14}_{-0.13}$, $0.53^{+0.05}_{-0.04}$, respectively. \cite{Yuasa10} used one partial covering absorber with $N_{\mathrm{H}}=114^{+34}_{-25}\times 10^{22} \,\mathrm{cm^{-2}}$ with a covering fraction of 0.36 in the 3-50 keV range which is consistent with our result. We noticed that in previous works with different observations, the X-ray spectra of XY Ari was modelled with from one to three partial covering absorber components. It may show that the absorption is variable over time and we cannot determine how much of the absorption comes from the accretion disc, vs. from the accretion curtains.

Many X-ray spectra of IPs indicate the presence of the fluorescent Fe $K\alpha$ line at 6.4 keV originating from the reflection of the hard X-rays from the WD surface \citep{Hellier04,Ezuka99}. The spectra of XY Ari show a prominent 6.4 keV fluorescence iron line feature assumed to be originating from the reflection of the hard X-ray emission from the WD surface. We find  that the equivalent width of the 6.4 keV line is about $51^{+12}_{-10}$ eV which is factor of two less than the previous determinations ($108^{+38}_{-38}$ eV, \cite{Salinas04}; $100^{+30}_{-90}$ eV, \cite{Ezuka99}).

The unabsorbed flux calculated in our study in the 2.0-10 keV band is 3.05 $\times 10^{-11} \, \mathrm{erg \,cm^{-2} s^{-1}}$ in accordance with the previous results indicating that the quiescent flux of XY Ari remains similar with small variations throughout the years. Our unabsorbed flux translates to a luminosity of $L_x=2.7^{+2.3}_{-1.6} \times 10^{32} \, \mathrm{erg \,s^{-1}}$ in the same energy range using the distance of \cite{Littlefair01} ($d= 270\pm100$ pc). If the 0.1-50 keV luminosity is assumed to be the accretion luminosity (e.g. ${L_x} \sim {L_{acc}}$), the mass accretion rate is calculated $\dot{M}=LR_{wd}/GM_{wd}=(2.3-3.8) \times 10^{-11} \, M_{\odot} yr^{-1}$ taking the mass of WD as $1.04 \, M_{\odot}$ \citep{Yuasa10} and the radius as $4.3-7.0 \times 10^{8}$ cm \citep{Hellier97}. 

The spin-phase resolved spectral analysis of XY Ari show change in the absorption columns for the photoelectric absorber (TBABS) and a partial covering absorber components over the spin phase. While the simple absorber (i.e. TBABS) accounts for the intrinsic absorption which is totally covering the source, the partial covering absorber accounts for the absorption column arising from the accretion curtain/column and all of them vary with the spin phase. The fits show increase of two absortion column at spin minima indicating the role of absorption. The NH values are highest at phases (0.1-0.3) and (0.6-1.0) that correspond to a hardening in HR (decrease) at spin phase (see Fig.6) which may be as a result of photoelectric absorption. In \textit{Chandra} observations from 2000, \cite{Salinas04} show a sinusoid-like behaviour of the absorption column as a function of spin phase. Our results confirm the variation of the absorption column paramters over the spin phase. The Analysis of the \textit{XMM-Newton} observation of XY Ari has also shown the pulse profile to be energy dependent. The energy-dependent nature of the X-ray spin pulse in XY Ari and the presence of absorption in its spectrum are seen generally in IPs indicating an accretion column absorbing structure. In most IPs,  spectral variations over the spin cycle and prominent phase-varying absorption are observed. Our results favour the absortion effects on the spin phase to be more dominant.

\label{lastpage}




\section*{Acknowledgements}

The authors thank an anonymous referee for the critical reading of the manuscript and for insightful remarks. DZC is supported by the 2219 scholarship programme of The Scientific and Technological Research Council of Turkey (TUBITAK) for this work on XY Ari. DZC gratefully acknowledges the warm hospitality of the the High-Energy Astrophysics Group at the Max Planck Institute for Extraterrestrial Physics during her visit. We acknowledge the use of public data from the \textit{XMM-Newton} data archive.

\bibliographystyle{mnras}
\bibliography{XYAri_MS_final} 



\end{document}